\documentclass[twocolumn,preprintnumbers,amsmath,amssymb,superscriptaddress,prl,aps]{revtex4-1}
\usepackage{graphicx}
\makeatletter

\newcommand{\Rmnum}[1]{\expandafter\@slowromancap\romannumeral #1@}
\makeatother

\begin{document}

\title{Structural versus electronic distortions of symmetry-broken IrTe$_2$}
\author{Hyo Sung Kim}
\affiliation{Center for Artificial Low Dimensional Electronic Systems, Institute for Basic Science (IBS), Pohang 790-784, Korea}
\affiliation{Department of Physics, Pohang University of Science and Technology, Pohang 790-784, Korea}
\author{Tae-Hwan Kim}
\affiliation{Center for Artificial Low Dimensional Electronic Systems, Institute for Basic Science (IBS), Pohang 790-784, Korea}
\affiliation{Department of Physics, Pohang University of Science and Technology, Pohang 790-784, Korea}
\author{Junjie Yang}
\affiliation{Laboratory for Pohang Emergent Materials, Pohang University of Science and Technology, Pohang 790-784, Korea}
\author{Sang-Wook Cheong}
\affiliation{Laboratory for Pohang Emergent Materials, Pohang University of Science and Technology, Pohang 790-784, Korea}
\affiliation{Rutgers Center for Emergent Materials and Department of Physics and Astronomy, Piscataway, New Jersey 08854, USA}
\author{Han Woong Yeom}
\email{yeom@postech.ac.kr}
\affiliation{Center for Artificial Low Dimensional Electronic Systems, Institute for Basic Science (IBS), Pohang 790-784, Korea}
\affiliation{Department of Physics, Pohang University of Science and Technology, Pohang 790-784, Korea}
\date{\today}

\begin{abstract}
We investigate atomic and electronic structures of the intriguing low temperature phase of IrTe$_2$ using high-resolution scanning tunneling microscopy and spectroscopy. We confirm various stripe superstructures such as $\times$3, $\times$5, and $\times$8. The strong vertical and lateral distortions of the lattice for the stripe structures are observed in agreement with recent calculations. The spatial modulations of electronic density of states are clearly identified as separated from the structural distortions. These structural and spectroscopic characteristics are not consistent with the charge-density wave and soliton lattice model proposed recently. Instead, we show that the Ir (Te) dimerization together with the Ir 5\textit{d} charge ordering can explain these superstructures, supporting the Ir dimerization mechanism of the phase transition.
\end{abstract}

\maketitle
\newpage

Strong spin-orbital coupling (SOC) has been widely recognized as the source of new physics in condensed matter systems such as various topological phases \cite{roy_topological_2009, kane_z_z_2_2005, moore_topological_2007, qi_topological_2008, konig_quantum_2007, bernevig_quantum_2006, hsieh_topological_2008, hsieh_majorana_2012, sato_topological_2010, fu_superconducting_2008, sau_generic_2010} and the SOC-induced Mott insulating state \cite{kim_novel_2008}. As a general strategy, it would be interesting to study how various quantum phases change under the influence of a strong SOC. In this respect, IrTe$_2$ is a very attractive materials. First of all, Ir has a strong SOC. It was shown to have a unique quasi 1D charge-density-wave-(CDW)-like ground state and the superconductivity emerges through the electron doping with a possibility of the quantum critical behavior \cite{yang_charge-orbital_2012}. It is natural to expect unusual behaviors in the CDW and superconducting states. Indeed, the recently observed superconductivity of Pt- or Pd-doped IrTe$_2$ was suspected as being non-conventional or topological \cite{pyon_superconductivity_2012, yang_charge-orbital_2012} but a very recent scanning tunneling spectroscopy (STS) study showed the conventional s-wave pairing behavior \cite{yu_fully_2014}. While further investigations are called for the nature of the superconductivity, the experimental evidence has been accumulated to indicate the unusual nature of the CDW-like phase transition ($T_c$~$\approx$~260~K without doping) with a wave vector of (1/5, 0, 1/5) (the $\times$5 phase, hereafter) \cite{machida_visualizing_2013}. Most importantly, the band gap opening does not occur but the strong restructuring of Fermi surfaces and band structures exists \cite{ootsuki_electronic_2013, fang_structural_2013, qian_structural_2013}. Extra intriguing aspects were also revealed such as the charge ordering in Ir 5$d$ orbitals \cite{pascut_dimerization-induced_2014}, the dimer-like distortions in both Ir and Te layers, and the depolymerization in the Te interlayer bondings \cite{cao_origin_2013, oh_anionic_2013}. At present, what is clear is the importance of the interplay of various different degrees of freedoms, structures for both Ir and Te layers, charges, orbitals, and bonds and to pin down the major driving force of this phase transition remains as a challenging task.

The charge-order or CDW nature of the ground state can best be checked by scanning tunneling microscopy and spectroscopy (STM/STS) studies since STM/STS can directly probe the spatial modulation of the local density of states (LDOS). Indeed, a very recent STM study identified various stripe superstructures of $\times$3, $\times$5, $\times$8, and $\times$11 \cite{hsu_hysteretic_2013} in topography. They explained this set of superstructures and the hysteresis of the transition in terms of 1D $\times$3 CDW ground state and the $\times$2 soliton excitation suggesting the low temperature phases as a rare example of the soliton lattice. However, this model is not easily reconciled with various recent electronic structure studies, which indicate the significant deviation from the CDW phase, and, more importantly, with most of the other works indicating the $\times$5 ground state. As discussed below, this discrepancy is largely related to the ambiguity of the STM topography in addressing complex charge and lattice orders entangled since STM is sensitive to both.

In this work, we performed a high resolution STM and STS study at 2, 4.3 and 78 K. While the previous STM study focussed only on the topography, we obtained both high resolution topography and detailed STS LDOS maps in order to extract the \textit{electronic} modulation of the superstructure. While we confirm the existence of various different superstructures, the LDOS and topographic characteristics of the stripe superstructure cannot be explained by soliton excitations. There is no evidence that the ground state is the $\times$3 CDW phase either. Instead, the structural and LDOS modulations can be understood based on the Ir (Te) dimerization and the charge ordering in Ir 5$d$. This result strongly suggests the Ir dimerization as the major driving force of the phase transition. We also emphasize that the direct interpretation of the STM topography as the electronic modulation is not generally justified.

The experiment was carried out with a commercial low-temperature STM (Specs, Germany). All topography measurements were performed in the constant-current mode with mechanically tapered Pt-Ir tips at mostly 78~K as cooled by liquid nitrogen but also at 4.3~K and 2~K as cooled by liquid He and the Joule-Thompson module, respectively. The differential conductance, $dI/dV$, was measured using the lock-in detection with a modulation of 1~kHz. The single crystals of IrTe$_2$ were grown by the Te flux using pre-sintered IrTe$_2$ polycrystals as reported before \cite{yang_charge-orbital_2012, hsu_hysteretic_2013, oh_anionic_2013}. The crystals were cleaved in ultrahigh vacuum at 86~K, which is much lower than $T_c$ \cite{yang_charge-orbital_2012}.

\begin{figure}
\includegraphics{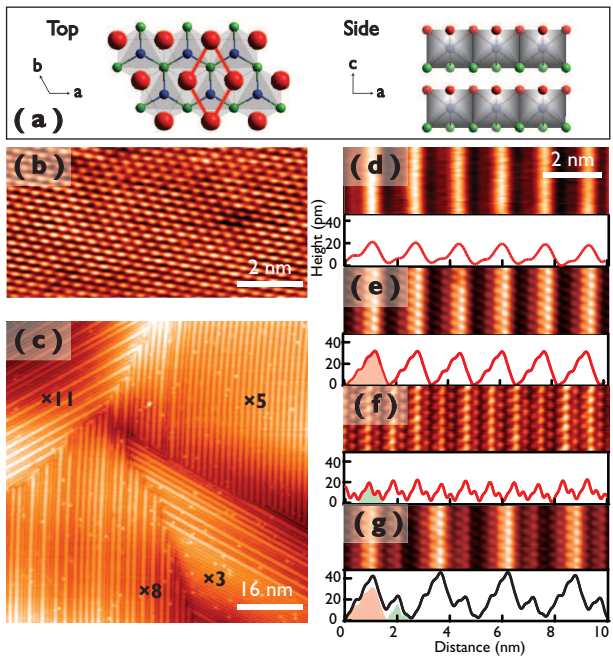}
\caption{\label{Fig1} (color online) (a) Structure model of undistorted IrTe$_2$. The upper Te, lower Te and Ir atoms are indicated by red, green, and blue balls, respectively. The in-plane unit cell is marked. (b) Room temperature STM image of Te terminated surface shows the undistorted 1$\times$1 structure. (c) Large scale STM topography showing domains of stripe phases (the sample bias of V$_s$ = 1 V). (d)--(g) Zoom-in STM topographies and the corresponding line profiles of three different stripe phases V$_s$ = 20~mV for (e), (f), and (g) and --0.5~V for (d); (d) the $\times$5 phase in filled states, (e) the same phase in empty states, (f) the $\times$3 and (g) $\times$8 phases in empty states. The building block structures of the line profiles are indicated by shaded areas. Line profiles are averaged along the wire direction.}
\end{figure}%

Figure~1 shows the structural model and the STM images of a cleaved surface of IrTe$_2$. A Te layer is exposed on the surface after cleaving since the polymeric bonds between Te layers is relatively weaker than the covalent bonding between Ir and Te layers \cite{jobic_occurrence_1992, jobic_anionic_1992}. The atomically resolved STM image [Fig.~1(b)] at RT above $T_c$ shows the undistorted Te surface layer with a three-fold symmetry. Below $T_c$, the strong stripe contrast appears in topography as a result of the symmetry-lowering phase transition. The three-fold symmetry of the surface dictates three equivalent orientations of the stripe domains as manifested by the intersecting domain boundaries in Fig.~1(c). Within each rotated domain, one can notice various different spacings between bright stripes. Sizable domains of regular spacings are formed with three particular spacings, $\times$3, $\times$5, and $\times$8 (3, 5, and 8 times of the lattice constant of IrTe$_2$, 1, 1.7, and 2.6~nm, respectively). While the $\times$5 domains are predominant at 78 K, the areal distribution of the $\times$3, $\times$5, and $\times$8 domains changes substantially not only upon the change of the temperature but also for different cleavages and for different parts of the surface. This result is more or less consistent with the previous STM study, which identified the same set of spacings, whose relative population changes systematically with the temperature \cite{hsu_hysteretic_2013}. In this work, we focus more on microscopic and spectroscopic structures of each regularly spaced domain letting aside the detailed temperature dependence of the domain distribution.

The detailed atomically resolved STM topographies for the stripe domains are shown in Figs. 1(d), 1(e), 1(f) and 1(g). As reported previously \cite{oh_anionic_2013}, the nonsinusoidal and asymmetric modulation of the $\times$5 phase [Figs.~1(d) and 1(e)] is characteristic. The $\times$3 phase has a much weaker, about half in the height scale, modulation of three Te rows [Figs. 1(f)]. While the recent study identified a marginal extra $\times$6 modulation within the $\times$3 phase as the alternation of the maximum height of the center row protrusion, our own measurement does not find any apparent ordering of such an extra modulation [Fig.~1(f)]. Thus, we simply call this structure as the $\times$3 phase. As clearly shown in Fig. 1(g), the $\times$8 (or larger) spacings can simply be understood to consist of one $\times$5 and single (or multiple) $\times$3 subunits or building blocks.

\begin{figure}
\includegraphics{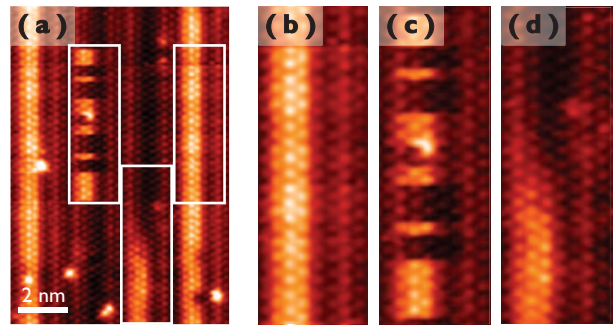}
\caption{\label{Fig2}(color online). Spatial and temporal fluctuations of stripes. (a) STM images (V$_s$ = 0.4~V) for four bright stripes within the typical $\times$8 phase. (b)-(d) Distinct types of bright stripes marked by rectangles in (a); (b) static, (c) dynamically fluctuating, and (d) spatially decaying ones.}
\end{figure}%

In contrast, the previous study postulated different building blocks, the $\times$3 CDW [Fig.~1(f) and the dark rows in Figs.~1(e) and 1(g)] and the $\times$2 soliton domain wall [the brightest rows in Figs.~1(e) and 1(g)]. This leads to conclude the $\times$3 phase, which is found to be the minority phase at 78 K, as the ground state and the $\times$5 as the excited state with a soliton lattice. We checked the same surface at 2 K as reached through the Joule-Thompson cooling module of our microscopy. However, we do not find any evidence that the $\times$3 phase dominates at this temperature ruling out the possibility of the $\times$3 CDW ground state. Moreover, the bias dependent STM images [see Figs.~1(d) and 1(e)] indicate that the characteristics of the CDW, the inversion of the filled and empty state images at low bias is absent. The protrusions (bright stripes) are persistent for a wide bias range of --0.5$\sim$1.0 eV indicating that the stripes in the STM topography are not due to a substantial electronic modulation. This is consistent with the lack of the band gap at the Fermi level discussed below, denying the CDW nature of the $\times$3 and $\times$5 phases.

The domain wall nature of the $\times$2 part (the bright rows) can also be ruled out. Figure~2 shows four particular bright stripes. While two such bright stripes are static, one in the center fluctuates between the bright and the dark contrast [Fig.~2(c)]. The other bright stripe statically fades into dark contrast [Fig.~2(d)]. If these stripes are soliton domain walls, their dynamic or static fluctuations must accompany the translation of neighboring domains, which is not observed at all. The CDW-soliton model is also not consistent with the STS result discussed below. Basically, the STS spectra indicate that the bright rows in topography do not have any unique electronic state in contrast to a soliton, which has a characteristic midgap state within the CDW gap. This unambiguously indicates that the bright rows are not due to an electronic excitation.

\begin{figure}
\includegraphics{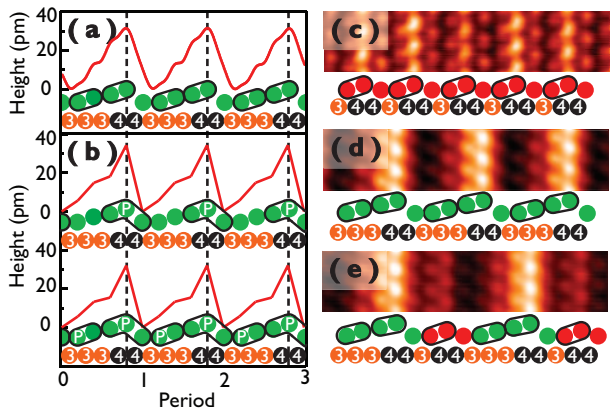}
\caption{\label{Fig3} (color online). (a) STM line profiles for the $\times$5 phase compared with (b) the height profile of the two structure models proposed; top from \cite{cao_origin_2013} and bottom from \cite{pascut_dimerization-induced_2014}. STM images ($I$~=~1~nA, V$_s$~=~20~mV) of (c) $\times$3, (d) $\times$5 (e) and $\times$8 phases with schematic models for the top Te (red and green circles) and Ir (orange and black circles) layers. The dimerization of Te atoms are marked by ovals with the their height variations and the charges of the Ir 5\textit{d} in the Ir layer (3+ and 4+) are indicated.}
\end{figure}%

The above discussion naturally leads to a structural origin for the stripe contrast as found in the dislocation network on the Au(111) surface \cite{narasimhan_elastic_1992}. Since STM topography picks up both the vertical lattice modulation and the LDOS modulation, one has to be careful to address either structural or electronic effect separately. Nevertheless, combined with detailed STS measurements, we may be able to choose a proper STM bias, normally a small bias near zero, where the LDOS modulation is featureless so that the height modulation can best be presented. We found +20 mV as such a bias condition in the present sample as shown below in more detail. The profiles in Figs. 1 and 3 correspond to those data. This height modulation for the $\times$5 phase agrees qualitatively well with the recent x-ray diffraction data \cite{yang_charge-orbital_2012, oh_anionic_2013} and the theoretical calculations \cite{cao_origin_2013, pascut_dimerization-induced_2014} as compared in the figure. That is, the bright rows correspond to Te dimers buckled up as pushed up by the Ir dimers underneath. The Te dimerization can further be confirmed by quantitatively measuring lateral Te-Te distances in STM images. The shorter Te-Te bonds are identified for the $\times$3, $\times$5, and $\times$8 structures, which are indicated as ovals in Fig.~3. The short bond lengths are between 260-300 pm contracted substantially from the long ones  by 10$\sim$20~\%. This dimerization is qualitatively consistent with the recent x-ray and first-principles calculation studies \cite{joseph_local_2013, cao_origin_2013, pascut_dimerization-induced_2014} and the quantitative difference between the theory and the present experiment can be due to the surface effect; the surface layer measured in the present STM experiment has different intralayer bondings, which were suggested not negligible \cite{hsu_hysteretic_2013}. The $\times$3 structure is formed by a single buckled dimer and the $\times$5 structure by two dimer units (see Fig.~3). The local structural origin of the stripes is well compatible with the fluctuation and the gradual decay of the stripes shown in Fig. 2.

While the stripe structures can be explained by the vertical and lateral distortion of the Te layer, which are closely related to the Ir dimerization underneath as the x-ray and the first-principles calculations suggested, one has to ask how this distortion affects the electronic structure and whether the electronic change is spatially uniform or modulated. In order to answer these questions, we took high resolution STS spectra at both RT and 4.3 K (Fig.~4). The DOS are strongly modified for a very wide energy range of more than $\pm$ 1 eV around the Fermi level across the phase transition. In particular the splitting of the broad spectral features at about 1.0, 0.2, and -0.7 eV  (rectangles) are drastic. The wide range DOS modification and the splitting or the formation of DOS dips are qualitatively consistent with the recent calculations but in sharp contrast to the CDW mechanism. That is, the Ir (Te) dimerization is shown to induce the splitting of Ir 5\textit{d} (Te 5\textit{p}) states \cite{pascut_dimerization-induced_2014} and the reduction of the Te 5\textit{p} DOS \cite{fang_structural_2013}.

\begin{figure}
\includegraphics{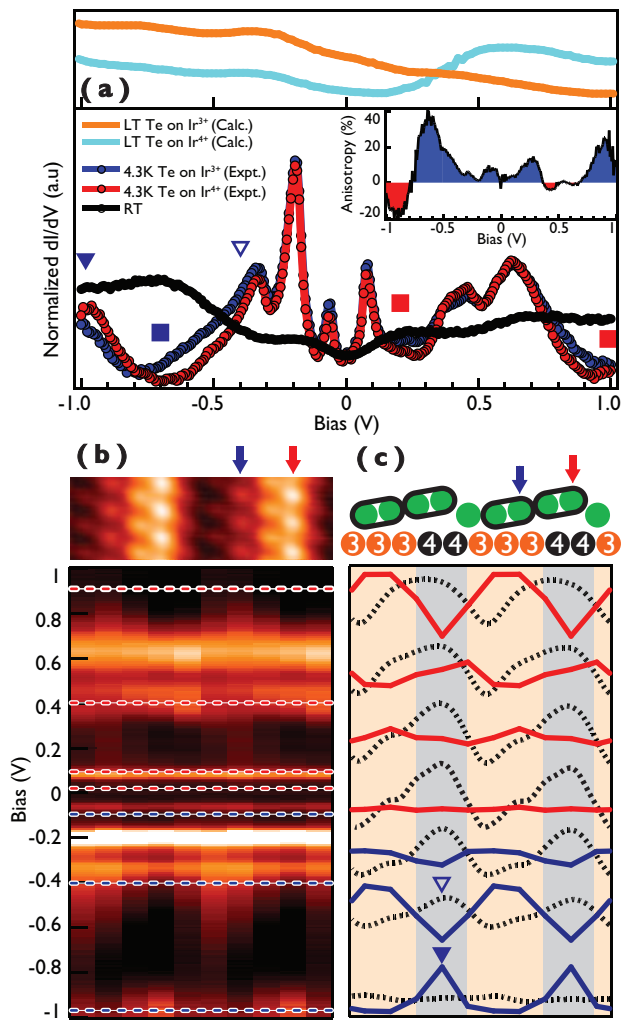}
\caption{\label{Fig4} (color online). (a) Averaged STS spectra for the room temperature phase (black dots) and for the bright (red) and dark (blue) contrast in  $\times$5 stripes at 4.3~K. The inset shows the LDOS difference between the bright and dark parts of a stripe. The spectral features of RT (squares) exhibit huge splittings at 4.3 K indicated by squares. Two line plots in the upper panel show the corresponding LDOS on two different Te atoms in the recent calculation \cite{pascut_dimerization-induced_2014}. (b) A STM topography and the STS ($dI/dV$ maps) spectra across two $\times$5 stripes obtained on the same positions. (c) The topographic profiles and the $dI/dV$ line scans [extracted from the $dI/dV$ maps in (b)]  for different biases indicated by dashed lines in (b) with the schematic structure model of the top Te and Ir layers. The grey parts correspond to the bright stripes and the buckled-up Te dimers.}
\end{figure}

We further mapped the spatial LDOS modulation, which would be directly related to the charge ordering. In the normalized $dI/dV$ map, the effect of the height modulation can be minimized to reflect the LDOS change dominantly \cite{feenstra_tunneling_1994, passoni_recovery_2009, koslowski_evaluation_2007}; the $dI/dV$ signal is normalized by tunneling current ($I/V$), which reflects largely the height variation. In Figs.~4(a) and 4(b), we can see that the LDOS spatial modulation is not large but is distinct [Fig.~4(b)] from the topographic ones (dashed lines) coming largely from the atomic height variation. In fact, the LDOS modulation is characteristically out of phase with the topography for a wide energy range of filled and empty states; the DOS are substantially reduced on the buckled-up Te atoms above the dimerized Ir atoms especially etween --0.8 and 0.3 eV and shows marginal enhancement between 0.3 and 0.7 eV (see the inset). This differentiates the two dimers within a single $\times$5 building block. This is in qualitative agreement with the recent calculation based on the Ir dimerization model while the calculation does not reproduce the fine details of STS spectra. However, the reduction of the LDOS on the buckled-up Te atoms above Ir dimers reflects the quintessential part of this model \cite{cao_origin_2013, pascut_dimerization-induced_2014}. The electronic modulation and the dimerization in the calculation were related to the charge ordering of Ir 5\textit{d}$^{3+}$ and Ir 5\textit{d}$^{4+}$ \cite{cao_origin_2013, pascut_dimerization-induced_2014} as depicted schematically in Figs.~3 and 4. Note also that the LDOS modulations are out of phase for the characteristic splittings mentioned above, especially the LDOS peaks at --1.0 and --0.4 eV [see the two bottom curves in Fig.~4(c)]. This suggests that the LDOS splitting is at least partly the bonding-antibonding type in accordance with the dimerization picture.

One important lesson of the present work is that the electronic modulations such as charge ordering or CDW need to be carefully addressed in STM studies since STM does not show the electronic modulation directly and the topography can largely be related to lattice distortions. Nevertheless, most of the previous STM studies for CDW \cite{dai_microscopic_2014} and charge order materials \cite{renner_atomic-scale_2002, carpinelli_surface_1997} interpreted topographies as representing electronic modulations. The limitation of the STM topography would be important for a systems with a substantial structural modulation as in the present case and more generally for diverse complex systems with various different degrees of freedom entangled.

In conclusion, the present high-resolution STM and STS study distinguishes the structural and electronic modulations of the intriguing low temperature stripe phases of IrTe$_2$. The structural and electronic contrasts are characteristically out of phase each other. The overall structural and electronic characteristics of the stripes are consistent with the Ir dimerization model but clearly rule out the CDW and soliton domain wall picture suggested previously. This study demonstrates clearly that the careful combination of STM and STS spectroscopic maps can unravel the entangled structural and electronic orders in complex systems such as charge order compounds.

This work was supported by Institute for Basic Science (Grant No. IBS-R014-D1). JJY and SWC were supported by the Max Planck POSTECH/KOREA Research Initiative Program [Grant No. 2011-0031558] through NRF of Korea funded by MEST. SWC was also supported by the NSF under Grant No. NSF-DMREF-1233349.

\end{document}